\documentclass[preprint,12pt]{elsarticle}
\usepackage{amssymb,amsmath}
\usepackage{a4wide,graphicx,amsthm,color}  
\usepackage{tabularx}  

\usepackage{dcolumn}
\usepackage{bm}

\def \FF {\mathfrak{F}}

\def \HH {\mathcal{H}}

\def \CCC {\mathbb{C}}

\def \RRR {\mathbb{R}}

\def \tr {\mathrm{Tr}}
\def \plus {\ + \ }

\def \equals {\ = \ }

\def \purity {\lambda_2}
\def \rms {\mathrm{s}}
\def \rmp {\mathrm{p}}

\newtheorem{proposition}{Proposition}

\newtheorem{definition}{Definition}

\begin{document}

\begin{frontmatter}
\title{Strongly separated pairs of core electrons in computed ground states of small molecules}

\author{Alex D. Gottlieb}
\address{Wolfgang Pauli Institute, Nordbergstrasse 15, 1090 Vienna, Austria}
\ead{alex@alexgottlieb.com}
\author{Rada M. Weish\"aupl}
\ead{rada.weishaeupl@univie.ac.at}
\address{Faculty of Mathematics, Vienna University, Nordbergstrasse 15, 1090 Vienna, Austria}

\date{\today}

\begin{abstract}
We have performed full configuration interaction computations of the ground states of the  molecules Be, BeH$_2$, Li, LiH, B, and BH and verified that the core electrons constitute ``separated electron pairs." 
%in the strong sense of the ansatz of antisymmetrized products of strongly orthogonal geminals (the APSG ansatz).   
These  separated pairs of core electrons have nontrivial structure; the core pair does not simply occupy a single spatial orbital.  

Our method of establishing the presence of separated electron pairs is direct and conclusive.  
We do not fit a separated pair model; we work with the wavefunctions of interest directly.  
To establish that a given group of spin-orbitals contains a quasi-separated pair, 
we verify by direct computation that the quantum state of the electrons that occupy those spin-orbitals is nearly a pure $2$-electron state. 
\end{abstract}


%\maketitle
 \end{frontmatter}

\section{Introduction}

According to separated pair theory \cite{Hurley,Surjan}, a many-electron wavefunction features an ideal ``separated electron pair" if and only if some subspace of the $1$-electron space, called the ``orbital domain" \cite{LiMaJiang} of the pair, is occupied by exactly two electrons in a pure state, i.e., a state described by a $2$-electron wavefunction or ``geminal."    Where this is approximately the case, we will say there is a ``quasi-separated electron pair."  

It has long been known \cite{AllenShull62,EbbingHenderson,MillerRuedenberg68} that the electronic structure of the Be atom and LiH molecule is very well described 
by the ansatz of anti-symmetrized products of strongly-orthogonal geminals, the APSG ansatz \cite{HurleyLennard-JonesPople,ParksParr,Kutzelnigg}.  
Though the APSG ansatz does not fit the BH molecule very well, the {\it core} electrons in BH can still be modeled as a separated pair.  
By 1970, Mehler, Ruedenberg, and Silver \cite{MehlerRuedenbergSilver} had determined that, ``The application of the separated pair approximation to Li, LiH, B, and BH has shown that the K [i.e., core] shell acts as a separated pair and is well described within the context of the separated pair approximation." 

The present study confirms this conclusion of  Mehler, Ruedenberg, and Silver 
\cite{MehlerRuedenbergSilver} in a direct and compelling way.   Ref. \cite{MehlerRuedenbergSilver} validates the separated pair structure {\it a posteriori}, by showing how well the separated pair model accounts for correlation energy and predicts other observables. 
In contrast, we discover the structure of the separated electron pairs by direct analysis of the wavefunction of interest, not its APSG approximation. 

We first obtain the wavefunction of interest at the full configuration interaction (FCI) level of theory.  
Given a subspace $G$ of the $1$-electron space, we derive, from the FCI wavefunction, the density matrix $D_G$ on the fermion Fock space over $G$ that  describes the quantum state of the electrons in $G$.  
If  $D_G$ is very close to a $2$-electron pure state, then $G$ may be regarded as the orbital domain of a quasi-separated electron pair. To quantify  how close $D_G$ is to a $2$-electron pure state, we shall use the ``$2$-purity" defined in Section~\ref{quasi}. 
% If $D_G$ represents a pure, $2$-electron state, then its $2$-purity equals $1$, but otherwise its $2$-purity is less than $1$.

In this way, we have found the orbital domains of the core electron pairs in the ground states of Li, LiH, Be, BeH$_2$, B, and BH.  
Each of these core pairs predominantly occupies a single spatial orbital $\phi_{core}$, in the sense that both spin-orbitals $\phi_{core \uparrow}$ and $\phi_{core \downarrow}$ are present in all of the dominant configurations 
of the FCI wavefunction.  The $2$-dimensional span of the spin-orbitals $\phi_{core \uparrow}$ and $\phi_{core \downarrow}$,  which we call the ``trivial core" orbital domain, may already be regarded as the orbital domain of a quasi-separated electron pair.  However, provided that one uses basis sets with enough core orbitals, one observes that the core pair's orbital domain involves several orbitals in addition to the trivial core orbitals.   To establish that those additional orbitals belong in the core orbital domain along with $\phi_{core \uparrow}$ and $\phi_{core \downarrow}$,  we will show that the state of the electrons in the larger orbital domain, including the additional orbitals, is even closer to a 2-electron pure state than is the state of the electrons in the trivial core.

%Features of the the structure of the core pairs can be predicted using separated pair theory, i.e., by variational minimization of the APSG ansatz energy.  

%Essentially, to establish the presence of a quasi-separated electron pair, we exhibit a subspace $G$ of the $1$-electron space such that $D_G$ is very close to a $2$-electron pure state, and we say that the subspace $G$ is an orbital domain of a quasi-separated pair.   
%According to this way of speaking, only an {\it ideal} separated pair will have a unique orbital domain, whereas a quasi-separated pair can be specified in a variety of ways, with a variety of possible orbital domains.
%When there are several competing orbital domains that describe essentially the same electron pair, the most suitable orbital domain among them is the one whose electrons are in the state of greatest $2$-purity.

Finding good orbital domains requires some guesswork.  There is a technique for locating possible orbital domains which has served us well for the cases treated in the present study, though it is not guaranteed to be effective in general.  We look for the kinds of statistical correlations that an ideal separated pair would entail, as discussed in Section~\ref{ideal}. Usage of this technique is shown by example in Section~\ref{the Be  subsection}, but it is not the main point  of emphasis of the present article.  

The main ideas of this article are that (i) it is feasible to check by direct computation whether a low-dimensional subspace of the $1$-electron space is the orbital domain of a quasi-separated electron pair,
and (ii) we have established in this way that the ground states of several small molecules actually feature quasi-separated core pairs.   

The rest of this article is organized as follows.  Section~\ref{separated} presents the theoretical background of our method, with a short history of separated pair theory in Section~\ref{brief}, definitions and properties of ideal separated electron pairs in Section~\ref{ideal}, and a more definitive description of our method in Section~\ref{quasi}.   Section~\ref{results} demonstrates our method by revealing the separated core pairs in  ground states of  Be, BeH$_2$, Li, LiH, and B and BH.   The Be atom is treated in the greatest detail in Section~\ref{the Be  subsection}.  There is also a brief conclusion (Section~\ref{conclusion}) and an appendix (\ref{marginal-normalized}).

\section{Separated electron pairs}
\label{separated}

\subsection{Brief history of separated pair theory}
\label{brief}

The separated electron pair approximation posits the ``antisymmetrized product of strongly orthogonal geminals" (APSG) ansatz for the wavefunction of the electrons in a molecule \cite{HurleyLennard-JonesPople,Kutzelnigg,Hurley}.  
The APSG ansatz for a $2n$-electron wavefunction $\Psi$ is 
\begin{equation}
\label{APSG}
      \Psi \equals \mathcal{A}_{2n} (g_1 \otimes g_2 \otimes \cdots \otimes g_n)\ ,
\end{equation}
where $\mathcal{A}_{2n}$ denotes an anti-symmetrizing operator and the $2$-electron wavefunctions or ``geminals"  $g_i(x_1,x_2)$ are such that 
\[
    \int dx_1 \overline{g_i(1,2)} g_j(1,2')   \equals {\bf 0} 
\]
for $i\ne j$.  The preceding condition of ``strong orthogonality" is equivalent \cite{Arai} to the condition that $g_i$ and $g_j$ be expressible as linear combinations of Slater determinants of spin-orbitals taken from mutually orthogonal subspaces $G_i$ and $G_j$.  

The APSG ansatz generalizes the Hartree-Fock ansatz, which results when all of the geminals in (\ref{APSG}) are taken to be $2$-electron Slater determinants.  
Thus the APSG ansatz variational energy is lower than the Hartree-Fock energy, and at one time it was hoped that the APSG ansatz might be broad enough to capture most of the correlation energy \cite{AllenShull61,AllenShull62}.  An early study of the beryllium iso-electronic series found that ``uniformly, about $90\%$ of the correlation energy is recovered" by the APSG ansatz \cite{MillerRuedenberg68}.  Unfortunately, subsequent studies found that the separated pair model ``works well for LiH but only partially for BH" \cite{MehlerRuedenbergSilver}, and demonstrated the ``uniform inability of the separated pair wavefunction to provide for correlation in the lone-pair geminal in NH" \cite{SilverRuedenbergMehler}.   It became clear that ``even an optimum separated-pair function can account for less than half the total correlation energy" in larger molecules \cite{Hurley}.  

Although APSG was not deemed successful as a simple variational ansatz, much work done within the last fifteen years shows that APSG wavefunctions perform very well as reference functions for correlated methods \cite{KallaySurjan,RostaSurjan, LiMaJiang, MaLiLi, SurjanSzabadosJeszneszkiZoboki}.     A series of four papers by Rassolov, Xu, and Garashchuck has shown that a viable ``geminal model chemistry" can be based on generalized separated pair theory \cite{Geminal Model Chemistry 1,Geminal Model Chemistry 2,Geminal Model Chemistry 3,Geminal Model Chemistry 4}.  

For more information and historical perspective, we refer the reader to the monograph by Hurley \cite{Hurley} that expounds pair theories as a conceptual bridge between Hartree-Fock theory and coupled cluster theory, and to a   
  review article by Surj\'an \cite{Surjan}, which devotes several pages to the history of geminal theories.
\subsection{Ideal separated electron pairs}
\label{ideal}

\subsubsection{Definitions for pure and mixed states}

McWeeny \cite{McWeeny} generalized the APSG ansatz to antisymmetrized products of strongly orthogonal group functions.   These are $N$-electron wavefunctions of the form 
\[
   \mathcal{A}_N \big[ f_1\otimes f_2 \otimes \cdots \otimes f_k \big]\ ,
\] 
% the operator $\mathcal{A}_N$ antisymmetrizes functions of $N$ variables, 
where each ``group function" $ f_i$ is an antisymmetric wavefunction of $n_i$ variables, $\sum n_i = N$, and the group functions are ``strongly orthogonal" in the sense that   
$$
    \int dx_1 \overline{f_i(1,2,\ldots,n_i)} f_j(1,2',\ldots,n'_j)   = {\bf 0}
$$
for $i\ne j$.
Strong orthogonality of the group functions $f_1,\ldots,f_k$ is equivalent \cite{Loewdin: little note} to the condition that the group functions $g_i$ can be expanded in Slater determinants taken from mutually orthogonal subspaces $G_i$ of the $1$-electron spin-orbital space.

Using this terminology, we can say that a many-electron wavefunction features an ``ideal separated electron pair" if it equals the antisymmetric product of two strongly orthogonal group functions, one of which is a geminal:
\begin{definition}
\label{pure} 
An $N$-electron wavefunction $\Psi$ features an ``ideal separated electron pair" if there exist strongly orthogonal group functions  $g(x_1,x_2)$ and $h(x_1,\ldots,x_{N-2})$ such that 
\begin{equation}
\label{ideal separated pair}
    \Psi=\mathcal{A}_N [g(1,2)h(3,\ldots,N)].
\end{equation}
\end{definition}

If $\Psi$ has the form (\ref{ideal separated pair}), then there exist orthogonal subspaces $G$ and $H$ of the $1$-electron space such that $g$ is a superposition of $2$-particle Slater determinants of spin-orbitals from $G$ while $h$ is a superposition of Slater determinants of $N-2$ spin-orbitals from $H$.  The subspace $G$ belonging to the separated electron pair has been called the pair's ``orbital domain" \cite{LiMaJiang}, ``Arai subspace" \cite{Surjan}, and ``carrier space" \cite{TomkachevDronskowski}.   

When an ideal separated electron pair with geminal $g$ has orbital domain $G$, we may say that $G$ is occupied by exactly two electrons in the pure state  $g$.  This is not just a figure of speech.  Whatever the $N$-electron wavefunction $\Psi$ may be, and whatever the subspace of the $1$-electron space $G$ may be, the ``open subsystem" \cite{GottliebHeadPerusse} consisting of just those electrons in $G$ is a many-electron system in its own right, in a quantum state that is determined by the state $\Psi$ of the entire molecule.  The quantum state of the electrons that happen to be in $G$ is typically a mixed state, prescribed formally by a density operator on the fermion Fock space over $G$.  In particular, there is typically a ``random" number of electrons in the subsystem, because electrons are free to move in and out of $G$.  But if $G$ is the orbital domain of an ideal separated electron pair, then there are {\it exactly} two electrons in $G$, and these two electrons are themselves in a pure $2$-electron state described by some geminal.   The converse statement holds as well, so that ideal separated electron pairs may be characterized as follows:

\begin{proposition}
\label{prop0}
A many-electron wavefunction $\Psi$ features an ideal separated electron pair with orbital domain $G$ if and only if the state of the many-electron system consisting of just those electrons in $G$, which is derived from the state $\Psi$ of the entire system, is a pure $2$-electron state.
\end{proposition}

Definition~\ref{pure} applies only to ``pure" states, those described by many-electron wavefunctions.   Proposition~\ref{prop0} now allows us to extend the notion of ideal separated electron pairs to ``mixed" many-electron states, which are described by density matrices:
\begin{definition}
\label{mixed}
A many-electron state features an ``ideal separated electron pair" if there exists a subspace $G$ of the $1$-electron space such that the derived state of the electrons in $G$ is a pure $2$-electron state.  The subspace G is called the ``orbital domain" of the separated electron pair.
\end{definition}

The formalism wherein the quantum state of the electrons in $G$ is derived from the electronic state of the entire molecule is elaborated below in Section~\ref{quasi}.

\subsubsection{Ideal separated electron pairs and natural spin-orbitals}
\label{the propositions}

Eigenvectors of the first-order density matrix (1RDM) derived from a many-electron wavefunction are called natural spin-orbitals (NSOs) associated with the wavefunction \cite{Loewdin}.   We shall use the L\"owdin normalization of the 1RDM so that its trace equals the number of electrons, instead of $1$.  Then the eigenvalue of the 1RDM corresponding to an NSO $\phi$ is called the natural occupation number of $\phi$.   We call NSOs of natural occupation number $0$ ``null" NSOs;  we are concerned only with non-null NSOs in the following.

Let $g$ denote a geminal wavefunction.  Then a non-null NSO associated with $g$ has even multiplicity as an eigenvector of the 1RDM derived from $g$. % \cite{Coleman}.  
Moreover, there exists  \cite{Coleman} some orthonormal sequence $\phi_1,\phi_2,\ldots$ of NSOs associated with $g$ such that 
\begin{equation}
\label{geminal}
     g \equals  c_1 | \phi_1 \phi_2 | \ + \  c_2 | \phi_3 \phi_4 | \ + \ c_3 | \phi_5 \phi_6 | \ + \ \cdots
\end{equation}

When a many-electron wavefunction $\Psi$ features an ideal separated electron pair with geminal $g$, all NSOs associated with $g$ are also NSOs associated with $\Psi$ with the same natural occupation number.  
Thus the orbital domain of an ideal separated electron pair is spanned by NSOs of the wavefunction.  
  
The preceding facts imply that the presence of an ideal separated electron pair entails certain statistical correlations between occupation events:

Let $\Psi$ be a many-electron wavefunction and let $\phi$ be a NSO associated with $\Psi$.  
By the ``multiplicity" of $\phi$ we mean its multiplicity as an eigenvector of the 1RDM derived from $\Psi$.  The following four propositions concern NSOs of multiplicity $2$.  
\begin{proposition}
\label{prop1}
Suppose that $\Psi$ features an ideal separated pair with orbital domain $G$.  
If $\phi$ is a non-null NSO of $\Psi$ of multiplicity $2$ then either $\phi \perp G$ or $\phi \in G$.  
\end{proposition}
\begin{proposition}
\label{prop2}
Suppose that $\phi_1$ and $\phi_2$ are non-null NSOs of $\Psi$ of multiplicity $2$, and suppose that 
$\phi_1 \in G$, the orbital domain of an ideal separated pair.  Then $\phi_1$ and $\phi_2$ have uncorrelated occupations if and only if $\phi_2 \perp G$. 
\end{proposition}
\begin{proposition}
\label{prop3}
 Suppose that $\phi_1$ and $\phi_2$ are non-null NSOs of $\Psi$ of multiplicity $2$ that both belong to the orbital domain of the same ideal separated pair.  If $\phi_1$ and $\phi_2$ have different natural occupation numbers then their occupations are perfectly anti-correlated, in the sense that they are never simultaneously occupied by electrons.  
 \end{proposition}
\begin{proposition}
\label{prop4}
 Suppose that $\phi_1$ is a non-null NSO of $\Psi$ of multiplicity $2$ that belongs to the orbital domain of an ideal separated pair.  If $\phi_2$ is an NSO of $\Psi$ that is orthogonal to $\phi_1$ but has the same natural occupation number,  then the occupations of $\phi_1$ and $\phi_2$ are perfectly correlated, in the sense that whenever one of the NSOs is occupied then the other is also occupied. 
\end{proposition}

To help us find quasi-separated pairs, we look for correlations between NSOs that approximate the ideal behavior described in the propositions.  
As we demonstrate in Section~\ref{the Be subsection}, correlations like those described in the above propositions are easily detected when we employ the ``marginal-normalized" correlation measure defined in \ref{marginal-normalized}.  

Correlations like those in the above propositions do not necessarily mean that a separated pair is present.  To establish conclusively that there is a separated pair, we proceed as described in the next section.

\subsection{Quantifying the purity of quasi-separated electron pairs}
\label{quasi}

A many-electron state features an ideal separated pair if and only if some subspace of the $1$-electron space is occupied by exactly two electrons in a pure $2$-electron state.   When this situation holds approximately, that is, when some subspace of the one $1$-electron space is occupied by {\it nearly} two electrons in a {\it nearly} pure state, we will say that the many-electron state features a ``quasi-separated" electron pair.   

In the preceding section we noted that the orbital domain of an {\it ideal} separated electron pair is necessarily spanned by natural orbitals.  
In contrast, an orbital domain of a quasi-separated electron pair need not be spanned by natural orbitals.
In Section~\ref{the BH subsection} we shall encounter an example where an orbital domain spanned by canonical Hartree-Fock orbitals is better (holds electrons in a state of greater purity)  than any orbital domain spanned by natural orbitals.  

The principal aim of the present article is to demonstrate the following method of determining whether a given many-electron state features a quasi-separated electron pair in a given orbital domain.
Given the many-electron state of interest as a correlated wavefunction or mixture of such, and given a subspace $G$ of the $1$-electron space, 
one derives the density matrix $D_G$ that describes the quantum state of the electrons in $G$.  
One can then assess how close $D_G$ is to the density matrix of a pure $2$-electron state and decide whether or not to regard $G$ as the orbital domain of a quasi-separated electron pair.  

We now explain in more detail how the density matrix $D_G$ is obtained, and how we quantify the proximity of $D_G$ to a pure $2$-electron state.

Let $G$ denote a subspace of the $1$-electron space 
$
\HH = L^2(\RRR^3 \times\{\alpha,\beta\}).
$
Slater determinants of $n$ spin-orbitals from $G$ represent configurations of $n$ electrons that all occupy $G$.  Let $G^{\wedge n}$ denote the span of all such Slater determinants.  
The fermion Fock space over $G$ is used to describe a system of finitely many electrons that all occupy $G$.  
This is the Hilbert space 
\begin{equation}
\label{Fock}
     \FF[G] \equals \CCC \oplus G \oplus  G^{\wedge 2}  \oplus  G^{\wedge 3} \oplus \cdots 
\end{equation}
whose component $G^{\wedge n}$ accommodates $n$-electron states.  
If $\hbox{dim}(G) = d$ is finite, then $\hbox{dim } \FF[G] = 2^d$ is also finite, because then the $d^{th}$ component of $\FF[G]$ is $1$-dimensional and all higher components are $0$-dimensional.

The system consisting of all the electrons in the molecule can be regarded as being of two parts: those electrons in $G$ and those in $G^{\perp}$, the ``orthogonal complement" of $G$.  Formally, this is expressed by the isomorphism \cite{AlickiFannesCapVI}
\begin{equation}
\label{isomorphism}
        \FF[G \oplus G^{\perp}] \ \cong  \ \FF[G]\otimes \FF[G^{\perp}] \ .
\end{equation}

The electronic state of a molecule determines the quantum state of its electrons that occupy $G$, as follows.  Suppose the electronic state of the entire molecule is represented by a density matrix $D$ on the Fock space $\FF[\HH] \cong \FF[G \oplus G^{\perp}]$.  Then the density matrix $D_G$ that represents the derived quantum state of the electrons in $G$ is the partial trace of $D$ over the tensor factor $\FF[G^{\perp}] $ in (\ref{isomorphism}), i.e., 
\begin{equation}
\label{substate density}
       D_G \equals  \tr_{\FF[G^{\perp}]} D\ .
\end{equation}
The density matrix $D_G$ is always block-diagonal with respect to the decomposition (\ref{Fock}) of $\FF[G]$, which we may express by writing 
\[
     D_G \equals D_{G,0}\oplus D_{G,1}\oplus D_{G,2}\oplus \cdots \oplus D_{G,\mathrm{dim}(G)}\ .
\] 
The probability that $G$ contains exactly two electrons equals $\tr D_{G,2}$.  In the event that two electrons are found in $G$, the state of those two electrons is given by the density matrix 
$\frac{1}{\tr D_{G,2}} D_{G,2}$.  Therefore, there are {\it exactly} two electrons in $G$ if and only if $\tr D_{G,2}= 1$, and those electrons are in a pure $2$-electron state if and only if 
 the largest eigenvalue of $D_{G,2}$ equals $1$.  

We denote the largest eigenvalue of $D_{G,2}$ by $\purity(G)$ and call it the ``$2$-purity" of the quantum state $D_G$.  
Thus $G$ is the orbital domain of an ideal separated electron pair if and only if $\purity(G) = 1$.  

We will use the $2$-purity of $D_G$ to judge whether or not $G$ may be regarded as the orbital domain of a quasi-separated electron pair.
Other measures of purity would serve the same purpose.   The standard measure of purity is the von Neumann entropy.
In fact, we originally used the von Neumann entropy to measure purity and were led to the same conclusions.
We prefer to report $2$-purity instead of von Neumann entropy because the former suits our purpose exactly: the $2$-purity of $D_G$ indicates whether $G$ contains nearly two electrons in a nearly pure state, whereas the von Neumann entropy of $D_G$ only quantifies purity and gives no information about particle number.  Also, $\purity(G)$ is somewhat easier to compute than the von Neumann entropy of $D_G$ because it only requires the $2$-particle component $D_{G,2}$.

\section{Results}
\label{results}

\subsection{Computational methods}

We used the electronic structure software GAMESS \cite{GAMESS} to obtain the full CI expansions of the wavefunctions of interest, using the natural orbitals as reference orbitals.  The electronic structures of the diatomic molecules  LiH, BH, and $\mathrm{Li_2}$ were computed with the experimental inter-nuclear distances 
$1.5957$\AA,  $1.2324$\AA,  and $2.6730$\AA, respectively; and the Be-H bond length in the linear molecule H-Be-H was set at $1.34$\AA\   \cite{NIST}.

The gaussian basis sets we used include the correlation consistent basis sets cc-pVDZ \cite{ccd} and their ``core-valence" extensions cc-pCVDZ \cite{WoonDunning}.   
We took the basis set cc-pCVDZ  for the Be atom from the EMSL Basis Set Exchange \cite{BSE1,BSE2}.    
We also used some even-tempered basis sets of gaussian primitives.   

FCI data obtained from GAMESS became the input for our own programs, which we used to derive the density matrices $D_G$ mentioned in Section~\ref{quasi}.    
With our current algorithms and computing power, computations begin to get quite heavy when $\hbox{dim}(G) \approx 20$.

\subsection{Nomenclature for natural orbital domains}
\label{nomenclature}

In Section~\ref{the data} we will identify the orbital domains of the separated core pairs in Be, BeH$_2$, Li, LiH, B, and BH.  This section describes the system of nomenclature we use for the orbital domains.

Consider the FCI/cc-pCVDZ ground state wavefunction for the Be atom.   
We have derived and diagonalized its 1RDM.  The eigenspaces of the 1RDM are invariant under the rotation group, and are in fact irreducibly invariant.    We call these eigenspaces of the 1RDM ``irreducible natural molecular shells." 
We label the irreducible natural molecular shells with s,p,d, etc., to signify which irreducible representation of the rotation group they support.  These eigenspaces are all even dimensional due to spin.    The $2$-dimensional eigenspaces are numbered 1s,2s,3s,4s, etc., in decreasing order of the corresponding eigenvalues of the 1RDM.   Similarly, the $6$-dimensional p shells are numbered 1p,2p,3p, etc., in decreasing order of average occupation.  Note that the most occupied natural p shell is labelled 1p.  The numeral $1$ in 1p is not an energy quantum number; it indexes occupation, so 1p in this context just means the most occupied irreducible natural p shell.

In Section~\ref{the Be subsection} we will establish that the FCI/cc-pCVDZ ground state for the Be atom features a quasi-separated pair of core electrons whose orbital domain is not merely the natural 1s shell, but also contains the irreducible natural 4s and 2p shells.   The orbital domain is the ``direct sum" of the natural 1s, 4s, and 2p shells.  It will be called the natural 1s$\oplus$4s$\oplus$2p shell.

The nomenclature for orbital domains is similar for the molecules of our study.  For example, consider a computed ground state wavefunction for the LiH molecule.  
We decompose the $1$-electron space into irreducible natural molecular shells by diagonalizing the 1RDM and classifying its eigenspaces by symmetry type $\sigma, \pi,\delta$, etc.  These irreducible natural molecular shells are then numbered 1$\sigma$,2$\sigma$,3$\sigma,\ldots$,1$\pi$,2$\pi$,3$\pi,\ldots$, etc., in order of decreasing average occupation number.   Having named the irreducible natural shells, larger natural molecular shells can be designated using direct sum notation.  For example, in Section~\ref{the LiH subsection} we will conclude that the FCI/cc-pCVDZ ground state of LiH features a quasi-separated electron pair whose orbital domain is the natural  1$\sigma\oplus$5$\sigma\oplus$2$\pi\oplus$6$\sigma$ shell.

The ground states of the Li and B atoms are degenerate. In these cases we have selected the symmetric mixed ground state for analysis.  Since the symmetric ground state is invariant under the rotation group, the eigenspaces of its 1RDM can also be classified by symmetry as 1s, 2s, 1p, 2p, etc.

\subsection{Data}
\label{the data}

\subsubsection{The Be atom}
\label{the Be subsection}

Quasi-separated electron pairs can be found in FCI ground states of the Be atom obtained using various basis sets.  
Here we report results for the basis sets cc-pVDZ and cc-pCVDZ, as well as an even-tempered basis set.  

First we present the results for the FCI/cc-pCVDZ ground state.  
Some of the data for this wavefunction is presented in Table~\ref{Be ccdc}.  
The natural occupation numbers of the NSOs are listed in the Occupation column.  
Note that these are the average occupations of the natural spin-orbitals, not the natural shells.  For example, 
to get the average occupation of the irreducible natural 1p shell, the spin-orbital occupation $0.02979$ in Table~\ref{Be ccdc} must be multiplied by $6$, since the 1p shell is spanned by six 1p NSOs.  

The main part of Table~\ref{Be ccdc} presents the values of the marginal-normalized correlation between NSO occupations.  
Marginal-normalized correlation,  denoted by $\kappa$, is defined in \ref{marginal-normalized}.  It is designed to detect the correlations expressed in Propositions~\ref{prop2}-\ref{prop4} of Section~\ref{ideal}.  
For uncorrelated occupations, as in Proposition~\ref{prop2}, the value of $\kappa$ is $0$.  For perfectly anti-correlated occupations, as in Proposition~\ref{prop3}, the value of $\kappa$ is $-1$.  
For perfectly correlated occupations, as in Proposition~\ref{prop4}, the value of $\kappa$ is $1$. 
Values of $\kappa$ are given to one decimal place.  An entry -$0$ in the table indicates that $\kappa$ is negative, and an entry ${\bf 1}$ or {\bf -}${\bf 1}$ in bold face indicates that $0.99 <  |\kappa | \le 1$.  

The values of $\kappa$ given in Table~\ref{Be ccdc} reveal the quasi-separated electron pair structure.   
The 2s, 1p, 3s, and 1d NSO occupations appear to be very highly correlated with one another but not with the 1s, 4s, and 2p NSO occupations.  
We therefore suppose -- and will soon confirm -- that the  2s, 1p, 3s, and 1d NSOs span the orbital domain of a quasi-separated electron pair, for this hypothesis is consistent with Propositions~\ref{prop1} - \ref{prop4}.    
The 1s, 4s, and 2p NSO occupations also appear to be highly correlated with one another, consistent with the hypothesis that they span the orbital domain of a quasi-separated electron pair.   
The 3p NSOs appear to be moderately correlated with NSOs from both presumptive separated pairs, but they are very lightly occupied and may be neglected.

\begin{table}
\small{\begin{tabular}{@{}c@{}c@{ } | @{ }r@{ }r@{ }r@{ }r@{ }r@{ }r@{ }r@{ }r@{ }r@{ }r@{ }r@{ }r@{ }r@{ }r@{ }r@{ }r@{ }r@{ }r}
     & Occupation   & 1s & 2s & 1p & 1p & 1p & 3s & 4s & 2p  & 2p & 2p & 1d & 1d & 1d & 1d & 1d  & 3p & 3p & 3p \\
     \hline
1s & 0.99855 & 0.9 & 0 & -0 & -0 & -0 & -0.1 & -0.9 & -1 & -1 & -1 & -0 & -0 & -0 & -0 & -0 & -0.5 & -0.5 & -0.5 \\ 
2s & 0.90854 & 0.1 & {\bf 1} & {\bf -1} & {\bf -1} & {\bf -1} & -0.9 & -0.1 & -0 & -0 & -0 & {\bf -1} & {\bf -1} & {\bf -1} & {\bf -1} & {\bf -1} & -0.7 & -0.7 & -0.7 \\ 
1p & 0.02979 & -0 & {\bf -1} & {\bf 1} & {\bf -1} & {\bf -1} & -0.9 & -0.1 & 0 & 0 & -0.1 & {\bf -1} & {\bf -1} & -1 & -1 & -1 & -0.8 & -0.6 & -0.8 \\ 
1p & 0.02979 & -0 & {\bf -1} & {\bf -1} & {\bf 1} & {\bf -1} & -0.9 & -0.1 & 0 & 0 & -0 & {\bf -1} & -1 & {\bf -1} & -1 & -1 & -0.8 & -0.8 & -0.6 \\ 
1p & 0.02979 & -0 & {\bf -1} & {\bf -1} & {\bf -1} & {\bf 1} & -0.9 & -0.1 & -0 & -0 & 0 & -1 & -1 & -1 & {\bf -1} & {\bf -1} & -0.6 & -0.8 & -0.8 \\ 
3s & 0.00184 & -0.1 & -0.9 & -0.9 & -0.9 & -0.9 & 0.9 & 0 & 0 & 0 & 0 & -0.9 & -0.9 & -0.9 & -0.9 & -0.9 & 0 & 0 & 0 \\ 
4s & 0.00052 & -0.9 & -0.1 & -0.1 & -0.1 & -0.1 & 0.3 & 0.7 & 0 & 0 & 0 & -0.1 & -0.1 & -0.1 & -0.1 & -0.1 & 0 & 0 & 0 \\ 
2p & 0.00027 & -0.9 & -0 & 0 & 0 & -0 & 0 & -0.4 & 0.9 & {\bf -1} & {\bf -1} & 0 & 0 & 0 & 0 & 0 & 0 & 0 & 0.1 \\ 
2p & 0.00027 & -0.9 & -0 & 0 & 0 & -0 & 0 & -0.4 & {\bf -1} & 0.9 & {\bf -1} & 0 & 0 & 0 & 0 & 0 & -0.4 & 0.1 & 0 \\ 
2p & 0.00027 & -0.9 & -0 & -0.1 & -0 & 0 & 0 & -0.4 & {\bf -1} & {\bf -1} & 0.9 & 0 & 0 & 0 & 0 & 0 & 0.1 & 0 & 0 \\ 
1d & 0.00007 & -0 & {\bf -1} & -0.9 & -0.9 & -0.8 & -0.9 & -0.1 & 0 & 0 & 0 & {\bf 1} & {\bf -1} & {\bf -1} & {\bf -1} & {\bf -1} & 0 & 0 & 0 \\ 
1d & 0.00007 & -0 & {\bf -1} & {\bf -1} & -0.8 & -0.8 & -0.9 & -0.1 & 0 & 0 & 0 & {\bf -1} & {\bf 1} & {\bf -1} & {\bf -1} & {\bf -1} & 0 & 0 & 0 \\ 
1d & 0.00007 & -0 & {\bf -1} & -0.8 & {\bf -1} & -0.8 & -0.9 & -0.1 & 0 & 0 & 0 & {\bf -1} & {\bf -1} & {\bf 1} & {\bf -1} & {\bf -1} & 0 & 0 & 0 \\ 
1d & 0.00007 & -0 & {\bf -1} & -0.8 & -0.8 & {\bf -1} & -0.9 & -0.1 & 0 & 0 & 0 & {\bf -1} & {\bf -1} & {\bf -1} & {\bf 1} & {\bf -1} & -0.3 & 0 & 0 \\ 
1d & 0.00007 & -0 & {\bf -1} & -0.8 & -0.8 & {\bf -1} & -0.9 & -0.1 & 0 & 0 & 0 & {\bf -1} & {\bf -1} & {\bf -1} & {\bf -1} & {\bf 1} & -0.3 & 0 & 0 \\ 
3p & 0.00001 & -0.4 & -0.4 & -0.7 & -0.8 & 0.3 & 0 & 0 & 0 & -0.2 & 0.2 & 0 & 0 & 0 & -0 & -0.1 & 0.3 & -0.9 & -0.9 \\ 
3p & 0.00001 & -0.4 & -0.4 & 0.3 & -0.9 & -0.7 & 0 & 0 & 0 & 0.1 & 0 & 0 & 0 & 0 & 0 & 0 & 0 & 0.3 & -0.9 \\ 
3p & 0.00001 & -0.4 & -0.4 & -0.9 & 0.3 & -0.8 & 0 & 0 & 0.1 & 0 & 0 & 0 & 0 & 0 & 0 & 0 & 0 & 0 & 0.3 \\ 
\end{tabular}}
\caption{
 Marginal-normalized correlations between NSO occupations of the FCI/cc-pCVDZ ground state of the Be atom.  
Correlations between NSOs of opposite spin appear on and below the main diagonal; correlations between NSOs of the same spin appear above the main diagonal.}
\label{Be ccdc}
\end{table}

We now argue that the core electrons form a quasi-separated pair whose orbital domain is the natural  1s$\oplus$4s$\oplus$2p shell.  
Of course, the natural 1s shell by itself may be also regarded as the orbital domain of a quasi-separated pair, for it is occupied by almost $2$ electrons in the pure Slater determinant state $|1\rms_\alpha 1\rms_\beta|$.    
However, including the 2p NSOs in the orbital domain of the core pair increases the purity of the state, and including the 4s NSOs as well increases the purity even further.  
We can quantify this increase in purity by measuring the ``$2$-purity" $\purity$ defined in Section~\ref{quasi}.  

Recall that  $\purity(G)$ denotes the largest eigenvalue of the $2$-particle component of the density matrix $D_G$ that represents the state of the electrons in $G$, so that $\purity(G)=1$ if and only if $G$ is the orbital domain of an ideal separated electron pair.  We observe that 
\begin{eqnarray*}
        \purity(1\rms) = 0.9985 \quad < \quad  \purity(1\rms\oplus 4\rms ) = 0.9987  & < &  \purity(1\rms\oplus 2\rmp) = 0.9992  \\
        & < &  \purity(1\rms\oplus 4\rms \oplus 2\rmp) = 0.9994 
        %         \purity(1\rms) = 0.99846 \quad < \quad \purity(1\rms\oplus 2\rmp) = 0.99917  \quad < \quad \purity(1\rms\oplus 4\rms \oplus 2\rmp) = 0.99944\ ,
\end{eqnarray*}
and conclude that the natural 1s$\oplus$4s$\oplus$2p shell is the orbital domain of a quasi-separated pair of core electrons.

Next we present the results for the smaller basis set cc-pVDZ.  The marginal-normalized correlations tabulated in Table~\ref{Be ccd} suggest that the 
2s$\oplus$1p$\oplus$3s$\oplus$1d shell may be the orbital domain of a quasi-separated pair.  Indeed it is, for its $2$-purity is $0.9999$. %$0.99993$.
The complementary shell, the natural 1s$\oplus$2p shell, has the same $2$-purity {\it a priori} and may also be regarded as the orbital domain of a quasi-separated pair.  

However, this core pair is virtually trivial: it does not amount to much more than a doubly occupied spatial 1s orbital.  The natural 1s$\oplus$2p shell has $2$-purity $0.9999$, but the trivial core orbital domain (the natural 1s shell by itself) actually has slightly higher $2$-purity.   Moreover, the data in Table \ref{Be ccd} does not bear  the signature of a distinctly separated core pair.   
According to Proposition~\ref{prop4}, if the 1s NSOs belonged to the orbital domain of an ideal separated pair of core electrons, then the marginal-normalized correlation $\kappa$ between the two 1s NSOs would equal $1$, but the actual value is only $0.4$.  
From Proposition~\ref{prop2}, we would also expect the 1s NSO and 2s NSO occupations to be nearly uncorrelated, but $\kappa=0.6$ between 1s and 2s  NSOs of opposite spin. 
The preceding considerations suggest that, with cc-pVDZ, the core electron pair is not distinctly separated.   

\begin{table}
\small{\begin{tabular}{c@{}c@{ } | @{ }r@{ }r@{ }r@{ }r@{ }r@{ }r@{ }r@{ }r@{ }r@{ }r@{ }r@{ }r@{ }r@{ }r}
     & Occupation   & 1s & 2s & 1p & 1p & 1p & 3s & 1d & 1d & 1d & 1d & 1d & 2p & 2p & 2p \\
     \hline
1s & 0.99996 & 0.4 & 0.1 & -0.1 & -0.1 & -0.1 & -0.1 & -0 & -0 & -0 & -0 & -0 & -0.6 & -0.6 & -0.6 \\ 
2s & 0.90798 & 0.6 & {\bf 1} & {\bf -1} & {\bf -1} & {\bf -1} & {\bf -1} & {\bf -1} & {\bf -1} & {\bf -1} & {\bf -1} & {\bf -1} & -0.4 & -0.4 & -0.4 \\ 
1p & 0.03010 & -0.1 & {\bf -1} & {\bf 1} & {\bf -1} & {\bf -1} & {\bf -1} & {\bf -1} & {\bf -1} & -1 & -1 & -1 & -0.6 & -0.1 & 0 \\ 
1p & 0.03010 & -0.1 & {\bf -1} & {\bf -1} & {\bf 1} & {\bf -1} & {\bf -1} & {\bf -1} & -1 & {\bf -1} & -1 & -1 & -0.6 & 0 & -0.1 \\ 
1p & 0.03010 & -0.1 & {\bf -1} & {\bf -1} & {\bf -1} & {\bf 1} & {\bf -1} & -1 & -1 & -1 & {\bf -1} & {\bf -1} & 0 & -0.6 & -0.6 \\ 
3s & 0.00138 & -0.1 & {\bf -1} & -1 & -1 & -1 & {\bf 1} & {\bf -1} & {\bf -1} & {\bf -1} & {\bf -1} & {\bf -1} & 0 & 0 & 0 \\ 
1d & 0.00007 & -0 & {\bf -1} & -1 & -1 & -0.9 & {\bf -1} & {\bf 1} & {\bf -1} & {\bf -1} & {\bf -1} & {\bf -1} & 0 & 0 & 0 \\ 
1d & 0.00007 & -0 & {\bf -1} & -1 & -0.9 & -0.9 & {\bf -1} & {\bf -1} & {\bf 1} & {\bf -1} & {\bf -1} & {\bf -1} & 0 & 0 & 0 \\ 
1d & 0.00007 & -0 & {\bf -1} & -0.9 & -1 & -0.9 & {\bf -1} & {\bf -1} & {\bf -1} & {\bf 1} & {\bf -1} & {\bf -1} & 0 & 0 & 0 \\ 
1d & 0.00007 & -0 & {\bf -1} & -0.9 & -0.9 & {\bf -1} & {\bf -1} & {\bf -1} & {\bf -1} & {\bf -1} & {\bf 1} & {\bf -1} & -0.3 & 0 & 0 \\ 
1d & 0.00007 & -0 & {\bf -1} & -0.9 & -0.9 & {\bf -1} & {\bf -1} & {\bf -1} & {\bf -1} & {\bf -1} & {\bf -1} & {\bf 1} & -0.6 & 0 & 0 \\ 
2p & 0.00001 & -0.4 & -0.6 & -0.7 & -0.7 & 0.5 & 0 & 0 & 0 & 0 & -0.1 & -0.5 & 0.5 & -0.9 & -0.9 \\ 
2p & 0.00001 & -0.4 & -0.6 & 0.2 & 0.3 & -0.7 & 0 & 0 & 0 & 0 & 0 & 0 & 0 & 0.5 & -0.9 \\ 
2p & 0.00001 & -0.4 & -0.6 & 0.3 & 0.2 & -0.7 & 0 & 0 & 0 & 0 & 0 & 0 & 0 & 0 & 0.5 \\ 
\end{tabular}}
\caption{
Marginal-normalized correlations between NSO occupations of the FCI/cc-pVDZ ground state of the Be atom.  
Correlations between NSOs of opposite spin appear on and below the main diagonal; correlations between NSOs of the same spin appear above the main diagonal.}
\label{Be ccd}
\end{table}

Apparently, the structure of the quasi-separated core pair is better revealed when we use the larger ``core-valence" basis set cc-pCVDZ. %rather than cc-pVDZ.  
We believe the reason is that the larger basis set includes more core orbitals and is therefore better able to reveal a core pair that is actually present in the ``exact" wavefunction, i.e.,  in the infinite basis set limit.   To test this, we have tried using large even-tempered basis sets.  These trials have repeatedly confirmed that the natural 1s$\oplus$4s$\oplus$2p shell deserves to be regarded as the orbital domain of a quasi-separated core pair.   

For example, here are the results for an even-tempered basis set with $10$ primitive gaussian s-orbitals and $4$ primitive gaussian p-orbitals.  We took the exponents of the primitives from Table~I of \cite{SchmidtRuedenberg}.  The exponents of the s-orbitals were those of the $N_s = 10$ series given for the Be atom, and the p-orbitals' exponents were 
those of the $N_p = 4$ series given for the B atom, scaled by
$
   \tfrac{16}{25} =  \tfrac{4^2}{5^2} 
$
to account for the main effect of the two atoms' differing nuclear charges.  
Table~\ref{Be even-tempered} shows the marginal-normalized correlations.  Again the correlations between the 1s and 2p NSO occupations stand out.   
Again, the $2$-purity of the natural 1s$\oplus$4s$\oplus$2p shell is greater than that of the natural 1s shell alone:
\begin{eqnarray*}
     \purity(1\rms) = 0.9981  \quad < \quad  \purity(1\rms\oplus 4\rms)  = 0.9982 & < &    \purity(1\rms \oplus 2\rmp)= 0.9989   \\
     & < & \purity(1\rms\oplus 4\rms \oplus 2\rmp) = 0.9990\ . 
    %    \purity(1\rms) = 0.9980868  \quad < \quad  \purity(1\rms\oplus 4\rms \oplus 2\rmp) = 0.9989947.  
\end{eqnarray*}
We tried several other even-tempered basis sets and got similar results.

\begin{table}
\small {\begin{tabular}{c@{}c@{ } | @{ }r@{ }r@{ }r@{ }r@{ }r@{ }r@{ }r@{ }r@{ }r@{ }r@{ }r@{ }r@{ }r@{ }r@{ }r@{ }r@{ }r }
     & Occupation   & 1s & 2s & 1p & 1p & 1p & 3s & 4s & 2p & 2p & 2p & 3p & 3p & 3p & 5s & 4p & 4p & 4p   \\
     \hline
 1s & 0.99830 & 0.9   & 0    & -0    & -0    & -0    & -0.1   & -0.8   & {\bf -1}   & {\bf -1}   & {\bf -1}   & -0.5   & -0.5   & -0.5   & -0.6   & -0.3   & -0.3   & -0.3      
\\ 2s & 0.90742 & 0.1  & {\bf 1}   & {\bf -1}   & {\bf -1}   & {\bf -1}   & -0.9   & -0.2   & -0.1   & -0.1   & -0.1   & -0.7   & -0.7   & -0.7   & -0.5   & -0.8   & -0.8   & -0.8   
\\ 1p &  0.03025  & -0    & {\bf -1}   & {\bf 1}   & {\bf -1}   & {\bf -1}   & -0.9   & -0.2   & -0.1   & -0    & 0    & -0.6   & -0.9   & -0.9   & -0.4   & 0    & -0.9   & -0.9     
\\ 1p &  0.03025  & -0    & {\bf -1}   & {\bf -1}   & {\bf 1}   & {\bf -1}   & -0.9   & -0.2   & -0.1   & 0    & -0    & -0.9   & -0.9   & -0.6   & -0.4   & -0.9   & -0.9   & 0      
\\ 1p & 0.03025   & -0    & {\bf -1}   & {\bf -1}   & {\bf -1}   & {\bf 1}   & -0.9   & -0.2   & 0    & -0.1   & -0.1   & -0.9   & -0.6   & -0.9   & -0.4   & -0.9   & 0    & -0.9      
\\ 3s & 0.00192 & -0.1   & -0.9   & -0.8   & -0.8   & -0.8   & 0.9   & 0    & 0    & 0    & 0    & 0    & 0    & 0    & -0.3   & 0    & 0    & 0   
\\ 4s & 0.00050  & -0.8   & -0.2   & -0.2   & -0.2   & -0.2   & 0.4   & 0.6   & 0    & 0    & 0    & 0    & 0    & 0    & 0    & -0.4   & -0.4   & -0.4  
\\ 2p &  0.00032  & -0.9   & -0.1   & -0    & -0.1   & 0.1   & 0    & 0    & 0.8   & {\bf -1}   & {\bf -1}   & 0    & 0.2   & 0    & 0    & -0.3   & 0    & -0.4  
\\ 2p &  0.00032  & -0.9   & -0.1   & 0    & 0.1   & 0    & 0    & 0    & {\bf -1}   & 0.8   & {\bf -1}   & 0    & 0    & 0.2   & 0    & -0.4   & -0.3   & 0     
\\ 2p &  0.00032  & -0.9   & -0.1   & 0.1   & 0    & -0.1   & 0    & 0    & {\bf -1}   & {\bf -1}   & 0.8   & 0.2   & 0    & 0    & 0    & 0    & -0.4   & -0.3   
\\ 3p &  0.00004  & -0.5   & -0.2   & 0.2   & -0.9   & -0.9   & 0    & 0    & 0    & 0    & 0.3   & 0.2   & -0.9   & -0.9   & 0    & 0.1   & 0    & 0      
\\ 3p &  0.00004  & -0.5   & -0.2   & -0.9   & -0.9   & 0.2   & 0    & 0    & 0.3   & 0    & 0    & -0.8   & 0.2   & -0.9   & 0    & 0    & 0.1   & 0      
\\ 3p &  0.00004  & -0.5   & -0.2   & -0.9   & 0.2   & -0.9   & 0    & 0    & 0    & 0.3   & 0    & -0.8   & -0.8   & 0.2   & 0    & 0    & 0    & 0.1      
\\ 5s & 0.00002 & -0.5   & -0.5   & -0.2   & -0.2   & -0.2   & 0.1   & 0.2   & 0    & 0    & 0    & 0    & 0    & 0    & 0.6   & 0    & 0    & 0    
\\ 4p &  0.00001  & -0.1   & -0.8   & 0    & -0.9   & -0.9   & 0    & 0    & 0    & 0    & 0.1   & 0.2   & 0    & 0    & 0    & 0.5   & -0.9   & -0.9      
\\ 4p & 0.00001   & -0.1   & -0.8   & -0.9   & -0.9   & 0    & 0    & 0    & 0.1   & 0    & 0    & 0    & 0.2   & 0    & 0    & -0.9   & 0.5   & -0.9    
\\ 4p & 0.00001   & -0.1   & -0.8   & -0.9   & 0    & -0.9   & 0    & 0    & 0    & 0.1   & 0    & 0    & 0    & 0.2   & 0    & -0.9   & -0.9   & 0.5    
\end{tabular}}
\caption{
Marginal-normalized correlations for the FCI ground state of the Be atom with an even-tempered basis set.  
Correlations between NSOs of opposite spin appear on and below the main diagonal; correlations between NSOs of the same spin appear above the main diagonal.
Ten NSOs of occupation less than $0.000005$ are omitted.  }
\label{Be even-tempered}
\end{table}

Our analysis shows that the Be atom ground state features a quasi-separated core pair in the natural 1s$\oplus$4s$\oplus$2p shell.   The separated core pair we observe appears to have the structure predicted by the separated pairs analysis of \cite{MillerRuedenberg68}, judging by the occupation numbers in Table XI of \cite{MillerRuedenberg68} or Table 2.6 in \cite{Hurley}. 

The character of the core pair has been evident since at least 1960, when  Linderberg and Shull \cite{LinderbergShull} reported that the inner pair of electrons in a $3$- or $4$-electron atom is very similar to the corresponding electron pair in the He iso-electronic series, while they judged correlation to be negligible between the shell of the core and the shell of the valence in Be.  They wrote, ``We may therefore conclude that the inner pair of electrons in these atoms [Li and Be] is very similar to the pair in the He-like series. This is a not unexpected result since a spherical shell of charge contributes nothing to the potential within the sphere, and the outer electrons in Li and Be-like atoms are largely in regions of space clearly outside the domain of the inner shell."

We have also investigated ground states of B$^+$, C$^{2+}$, and N$^{3+}$, which are iso-electronic with Be.  When  we use cc-pCVDZ or large even-tempered basis sets to compute them, the ground states of these ions are generally found to possess quasi-separated core pairs in their natural 1s$\oplus$2p shells.  By comparing $2$-purities, we can clearly see that the orbitals homologous to the natural 4s orbitals in the Be core do not belong in the   C$^{2+}$ andd N$^{3+}$ core orbital domains.   In B$^+$, however, there is a case to be made for inclusion of the 4s NSOs in the core domain. Using cc-pVDZ, the $2$-purities do increase slightly when the 4s NSOs are included:
\begin{eqnarray*}
     \purity(1\rms) = 0.99903  \quad < \quad  \purity(1\rms\oplus 4\rms)  = 0.99906 & < &    \purity(1\rms \oplus 2\rmp)= 0.99947 \\
     & < & \purity(1\rms\oplus 4\rms \oplus 2\rmp) = 0.99949\ . 
%     \purity(1\rms) = 0.99903  \quad < \quad  \purity(1\rms\oplus 4\rms)  = 0.9990597 & < &    \purity(1\rms \oplus 2\rmp)= 0.9994684 \\
%     & < & \purity(1\rms\oplus 4\rms \oplus 2\rmp) = 0.9994949\ . 
\end{eqnarray*}
On the other hand, a slightly negative effect of including the 4s orbitals is observed when we use the von Neumann entropy $S(D_G) = -\tr(D_G\log_2 D_G)$ to measure the purity of $D_G$.  
Including the 4s NSOs in the orbital domain $G$ increases the entropy (and thereby decreases the purity) of $D_G$:
\begin{eqnarray*}
     S(D_{1\rms}) = 0.0118  \quad < \quad  S(D_{1\rms\oplus 4\rms})  = 0.0129 & < &    S(D_{1\rms \oplus 2\rmp})= 0.0078 \\
     & < & S(D_{1\rms\oplus 4\rms \oplus 2\rmp}) = 0.0083\ . 
%S(1\rms) = 0.01180131  \quad < \quad  S(1\rms\oplus 4\rms)  = 0.01293267 & < &    S(1\rms \oplus 2\rmp)= 0.007796974 \\
 %   & < & S(1\rms\oplus 4\rms \oplus 2\rmp) = 0.008317359\ . 
\end{eqnarray*}

Thus, the 4s NSOs appear to evaporate out of the core pair's orbital domain as we move up the Be-isoelectronic series from Be to N$^{3+}$.  
An anonymous referee remarks that this phenomenon probably reflects the fact \cite{Davis,Gimarc,BanyardBaker,SahaEtAl} that the ratio of ``radial correlation" to ``angular correlation" \cite{TaylorParr} decreases along the He iso-electronic series, for the core electron pair in a Be-isoelectronic atom is expected to behave much like the two electrons in the corresponding He-isoelectronic atom.

\subsubsection{{\rm BeH$_{\mathrm 2}$ } }

The FCI/cc-pCVDZ ground state of BeH$_2$ features a quasi-separated pair of core electrons in its natural 1$\sigma_g\oplus$5$\sigma_g\oplus$4$\sigma_u\oplus$3$\pi_u$ shell:
\[
%    \purity(1\sigma) = 0.99839 \quad < \quad \purity(1\sigma_g\oplus 5\sigma_g\oplus 4\sigma_u\oplus 3\pi_u) = 0.99922.
   \purity(1\sigma_g) = 0.9984 \quad < \quad \purity(1\sigma_g\oplus 5\sigma_g\oplus 4\sigma_u\oplus 3\pi_u) = 0.9992.
\]  
The 1$\sigma_g\oplus$5$\sigma_g\oplus$4$\sigma_u\oplus$3$\pi_u$ shell of the core electrons in BeH$_2$ corresponds closely to the 1s$\oplus$4s$\oplus$2p shell of the core electrons in the Be atom, with the 2p shell of Be splitting into the 4$\sigma_u$ and 3$\pi_u$ shells of BeH$_2$.

\subsubsection{The {\rm Li} atom }
\label{the Li subsection}

The ground state of the Li atom is degenerate $^2$S.   The ground state subspace is spanned by two wavefunctions $\Psi_\alpha$ and $\Psi_\beta$ of opposite spin.  We consider the mixed ground state  
$    \tfrac12 |\Psi_\alpha\rangle\!\langle\Psi_\alpha | + \tfrac12 |\Psi_\beta \rangle\!\langle\Psi_\beta |  $ and label the eigenspaces of its 1RDM as described in Section~\ref{nomenclature}.  

In the FCI/cc-pCVDZ ground state,
\begin{eqnarray*}
%    \purity(1\rms) = 0.99714 \quad < \quad \purity(1\rms\oplus 3\rms\oplus 1\rmp) = 0.99972.
   \purity(1\rms) = 0.9971 \quad < \quad \purity( 1\rms\oplus 3\rms ) = 0.9983 & < & \purity(1\rms\oplus 1\rmp) = 0.9985 \\
      & < & \purity(1\rms\oplus 3\rms\oplus 1\rmp) = 0.9997.
\end{eqnarray*}
With the smaller basis set cc-pVDZ, the trivial core has the greatest purity, with $\purity(1\rms)=0.9999$. %0.99992$.
However, when the basis set cc-pVDZ is de-contracted, and we make a basis set out of all of its primitives, the natural 1s$\oplus$3s$\oplus$1p shell reemerges as the orbital domain of the core.

\subsubsection{{\rm LiH}}
\label{the LiH subsection}

The FCI/cc-pCVDZ ground state of LiH features a quasi-separated core pair located in the natural 1$\sigma\oplus$5$\sigma\oplus$2$\pi\oplus$6$\sigma$ shell: 
\[
%     \purity(1\sigma) =  0.99713  \quad < \quad \purity(1\sigma\oplus 5\sigma\oplus 2\pi\oplus 6\sigma) = 0.99937.
     \purity(1\sigma) =  0.9971  \quad < \quad \purity(1\sigma\oplus 5\sigma\oplus 2\pi\oplus 6\sigma) = 0.9994.
\]
This orbital domain is closely related to the orbital domain of the core electrons in the Li atom, the natural 1s$\oplus$3s$\oplus$1p shell: the natural 1p shell in Li splits into the 2$\pi$ and 6$\sigma$ shells in LiH.  

%It was too difficult to compute the density matrix that describes the state of the electrons in the shell complementary to the natural 
%1$\sigma\oplus$5$\sigma\oplus$2$\pi\oplus$6$\sigma\oplus$9$\sigma$ shell, because its dimension is too large, but we know {\it a priori} that it has the same $2$-purity. 
%We were able to compute the density matrix for the natural 2$\sigma\oplus$3$\sigma\oplus$1$\pi\oplus$4$\sigma\oplus$7$\sigma\oplus$3$\pi\oplus$8$\sigma$
%shell, and found that its $2$-purity is $0.99936$.

We do not detect an interesting quasi-separated core pair when we use the smaller basis set cc-pVDZ.
The orbital domain of the quasi-separated core pair in the FCI/cc-pVDZ ground state of LiH amounts to little more than the natural 1$\sigma$ shell, which already has $2$-purity 
$0.9999$.  
Including other orbitals brings the average occupation closer to $2$ electrons but does not increase purity.
However, when we use large even-tempered basis sets, we again discover a quasi-separated core pair with orbital domain 1$\sigma\oplus$5$\sigma\oplus$2$\pi\oplus$6$\sigma$.

% Rada's basis set for LiH is an even-tempered basis consisting of s and p atomic orbitals on the Li and H atoms.  
%The Li and H s-orbitals are those of in Table~I of \cite{SchmidtRuedenberg} with $N_s = 16$ and $N_s = 4$, respectively, 
% the Li p-orbitals have exponents $0.0025 4^k, k=1,\ldots,6,$ and H has one p-orbital with exponent $0.727$

\subsubsection{The {\rm B} atom }
\label{the B subsection}

The ground state of the B atom is $^2P$, so the ground state subspace is $6$-dimensional, spanned by six orthonormal wavefunctions $\Psi_\alpha^x, \Psi_\alpha^y,\Psi_\alpha^z, \Psi_\beta^x, \Psi_\beta^y,\Psi_\beta^z$.  
We choose a definite ground state to work with.  Any wavefunction in $\hbox{span}\{ \Psi_\alpha^x,\ldots ,\Psi_\beta^z\}$ represents a pure ground state, and any density matrix that is spectrally supported on $\hbox{span}\{ \Psi_\alpha^x,\ldots ,\Psi_\beta^z\}$ represents a mixed ground state.
Our choice is the symmetric mixed state 
\begin{equation}
\label{symmetrixed}
  \tfrac16 |\Psi_\alpha^x \rangle\!\langle \Psi_\alpha^x |  \plus  \tfrac16 |\Psi_\alpha^y \rangle\!\langle \Psi_\alpha^y |  
  \plus  \tfrac16 |\Psi_\alpha^z \rangle\!\langle \Psi_\alpha^z | 
  \plus  \tfrac16 |\Psi_\beta^x \rangle\!\langle \Psi_\beta^x | 
  \plus  \tfrac16 |\Psi_\beta^y \rangle\!\langle \Psi_\beta^y | 
  \plus  \tfrac16 |\Psi_\beta^z \rangle\!\langle \Psi_\beta^z  |\ .
\end{equation}
This state is invariant under the group action, so we may name the irreducible natural shells just as described in Section~\ref{nomenclature}.

In the FCI/cc-pCVDZ ground state of the form (\ref{symmetrixed}), 
\begin{eqnarray*}
     \purity(1\rms) =  0.9990 % 0.999003
            \quad < \quad 
       \purity(1\rms \oplus  4\rms) = 0.9992 % 0.9992325 
       & < & 
       \purity(1\rms \oplus  3\rmp) =  0.9994  % 0.9993829 
       \\
       & < & 
       \purity(1\rms \oplus 3\rmp \oplus 4\rms) = 0.9996 \ .% 0.9996133
\end{eqnarray*}
We conclude that the symmetric mixed state  (\ref{symmetrixed}) features a quasi-separated core pair in its natural 1s$\oplus$3p$\oplus$4s shell.

\subsubsection{{\rm BH} }
\label{the BH subsection}

The FCI/cc-pCVDZ ground state of BH features a quasi-separated pair of core electrons in its natural 1$\sigma\oplus$4$\pi\oplus$10$\sigma$ shell:
\[
%    0.99898 <  0.9991339 <  0.99920.
   \purity(1\sigma) = 0.9990  \quad < \quad \purity(1\sigma\oplus 4\pi ) = 0.9991 \quad < \quad \purity(1\sigma\oplus 4\pi \oplus 10\sigma) = 0.9992.
\]  

As mentioned in Section~\ref{quasi}, the ``best" orbital domain of a quasi-separated electron pair need not be a natural molecular shell.   
For the FCI/cc-pCVDZ ground state of BH, the best orbital domain we have found is a Hartree-Fock (HF) molecular shell, spanned by canonical HF orbitals.   
This orbital domain $K$ is the direct sum of three irreducible HF $\sigma$ shells and one irreducible HF $\pi$ shell. 
The $2$-purity of the state $D_K$ is $0.9996$.  

The orbital domain $K$ would seem to be a slight perturbation of the orbital domain found for the core pair of the B atom in the preceding section.   
We have not computed the overlaps, but occupation numbers tell the story.  
The irreducible $\sigma$ shells in $K$ have average occupations $1.99812$, $0.00067$, and $0.00032$, and the $\pi$ shell has average occupation $0.00065$.  
The $\sigma$ shells with average occupation $1.99812$ and $0.00067$ correspond to the natural 1s and 4s shells of the symmetric mixed state (\ref{symmetrixed}) of the B atom, which have average occupations $1.99817$ and $0.00065$, respectively.   The irreducible natural 3p shell in the orbital domain of the B atom core seems to have split into the $\pi$ shell of average occupation $0.00065$ and the $\sigma$ shell of average occupation $0.00032$, for $0.00065 + 0.00032 = 0.00097$ is very close to the average occupation of the B atom's natural 3p shell, i.e., $0.00098$.  
This is all consistent with the results of the separated pair analysis of \cite{MehlerRuedenbergSilver} concerning the ``transferability of the K [core]  shell pair in B and BH."

We do not find a nontrivial core orbital domain when we use the smaller basis set cc-pVDZ.  However, when this basis set is de-contracted, and we use the basis set consisting of all the primitive gaussian orbitals that contribute to cc-pVDZ, we do find that 
\[
   \purity(1\sigma) = 0.9987   \quad < \quad \purity(1\sigma\oplus 4\pi \oplus 10\sigma) = 0.9988  \quad < \quad \purity(1\sigma\oplus 4\pi ) = 0.9989\ .  
\]

\section{Conclusion}
\label{conclusion}

The presence of quasi-separated electron pairs can be established conclusively by direct computation when a sufficiently correlated representation of the many-electron state of interest is available.  
It is convenient to work with the wavefunction in CI form, but it need not be the {\it full} CI wavefunction.    We limited the present study to FCI wavefunctions in order that our specific conclusions concerning the ground states of Li, LiH, Be, BeH$_2$, B, and BH would be as compelling as possible.   In future work, we may apply our method to larger molecules by working with coupled cluster wavefunctions.

% A practical limitation of our method is that it takes too much time to compute the quantum state of the electrons in a presumptive orbital domain $G$ if $\dim(G)$ is too large.   

Using our method, we have shown that FCI ground states of Li, LiH, Be, BeH$_2$, B, and BH all appear to feature interesting quasi-separated pairs of core electrons, provided large enough basis sets are used.  
In each case, we have identified the main components of the core pair's orbital domain.   Our results are consistent with the predictions of separated pair theory \cite{Hurley,MillerRuedenberg68, MehlerRuedenbergSilver}.

% To find the orbital domains of the quasi-separated pairs, we looked for the kind of statistical correlations that are described in the propositions at the end of Section~\ref{ideal}.  
% As demonstrated with the Be atom in Section~\ref{the Be subsection}, these correlations were easy to detect when we used the measure of ``marginal-normalized correlation" defined in the appendix.  
% Possible orbital domains were tested by deriving the quantum states of the electrons that occupy them and computing the $2$-purities of these states, as discussed in Section~\ref{quasi}.    

% The striking results of the present study suggest that core electrons generally tend to form separated electron pairs, which can  be ``factored out" of the many-electron wavefunction as strongly orthogonal geminals.  

\bigskip

\section*{Acknowledgements}
%\begin{acknowledgments}
 A.G. was supported by the Vienna Science and Technology Fund (WWTF) project MA07-037 on ``Correlation in Quantum Systems."  R.W. is supported by the Hertha Firnberg Program, funded
by the Austrian Science Fund (FWF) project T402-N13.   We thank John Head for a great deal of good advice and helpful guidance, without which we would not have accomplished this work.  We thank Robert Hammerling for his physical insight.  
%\end{acknowledgments}

\appendix
\section{Marginal-normalized correlation}
\label{marginal-normalized}

In this section we define the ``marginal-normalized correlation" $\kappa$ that we used in Section~\ref{the Be subsection} to test for statistical correlations like those in Propositions~\ref{prop2} - \ref{prop4} of Section~\ref{ideal}.

Suppose that $\phi_1$ and $\phi_2$ are two NSOs associated to a many-electron state, and let $G = \hbox{span}\{\phi_1,\phi_2\}$.  The quantum state $D_G$ of the electrons in $G$ is characterized by three numbers, namely, the probability $n_1$ that $\phi_1$ is occupied, the probability $n_2$ that $\phi_2$ is occupied, and the probability $p$ that $\phi_1$ and $\phi_2$ are occupied simultaneously.   The probability that $\phi_1$ and $\phi_2$ are both unoccupied is then $1 - n_1 - n_2 +p$, and it can be shown that there are no ``off-diagonal" correlations.   The marginal-normalized correlation between two NSOs is simply a function of $n_1,n_2$, and $p$.

Let us examine the statistical correlation between the occupations of two NSOs from our data set.  We have selected two NSOs 
whose occupation probabilities are $0.9288$ and $0.0010$, rounded to $4$ decimal places.  
The probability that both NSOs are occupied simultaneously is $0.0003$.   The joint occupation probabilities can be arranged in a $2\times 2$ table: 
\begin{equation}
\begin{tabular}{rccc|c }
                  &  \qquad &    vacant  &      occupied   &  \\
  vacant  &  & 0.0705 & 0.0007  &  0.0712 \\
    occupied &  & 0.9285 & 0.0003  & 0.9288 \\
    \hline
                    & &  0.9990  &  0.0010 &  \\
\end{tabular}
\label{first table}
\end{equation}
The numbers on the right and bottom are the marginal sums, which are the occupation and vacancy probabilities for the two NSOs.

From the preceding probabilities, we can compute measures of correlation.  We may assign numerical values to ``vacancy" and ``occupation" and compute the familiar (Pearson) correlation coefficient of the resulting random variables (it does not matter which numerical values one chooses; the correlation coefficient will come out the same because there are only two events: occupation and vacancy).  For the probabilities in table~(\ref{first table}) above, $\rho \approx -0.077$.

The correlation coefficient $\rho$ does not suit our purposes because it does not take the marginal probabilities into account.    
For example, a table of joint probabilities with the same marginals as table (\ref{first table}) above can only have $\rho$ between $-0.114$ and $0.009$ (rounding to $3$ places).  These extremes are realized by the tables 
\begin{equation}
\begin{tabular}{ cc|c }
   0.0702 & 0.0010  &  0.0712 \\
    0.9288 & 0.0000  & 0.9288 \\
    \hline
   0.9990  &  0.0010 &   \\
\end{tabular}\qquad\qquad \hbox{ and } \qquad\qquad
\begin{tabular}{ cc|c }
  0.0712 & 0.0000  &  0.0712 \\
  0.9278 & 0.0010  & 0.9288 \\
    \hline
     0.9990  &  0.0010 &   \\
\end{tabular}
\label{second and third tables}
\end{equation}
We prefer to use a measure of correlation that depends on the marginals, but which can attain a minimum of $-1$ and a maximum of $+1$.  
The variable $p$ in a table of probabilities 
\medskip
\begin{center}
\begin{tabular}{ ccc|c }
 $ 1 - n_1 - n_2 +p$  & \quad\quad & $n_2 - p$ &  $1 - n_1$ \\
 $ n_1-p $ &   & $p$   & $n_1$ \\
    \hline
   $  1-n_2 $  &   & $ n_2 $ & $ $  \\
\end{tabular}
\end{center}
\medskip
\noindent can range between $\max\{0, n_1+n_2 - 1\}$ and $\min\{n_1,n_2\}$, and independence occurs when $p = n_1n_2$. 
The greatest that $|p - n_1n_2|$ can be is 
\begin{eqnarray*}
    \min\{n_1,n_2\} - n_1n_2 \qquad  &\hbox{ if }& \quad p \ge n_1n_2 \\
       n_1n_2 -\max\{0, n_1+n_2 - 1\} \qquad   &\hbox{ if }& \quad p < n_1n_2 \ ;
\end{eqnarray*}
accordingly, we define the marginal-normalized correlation $\kappa(n_1,n_2,p)$ to be  
\[
    \frac{p-n_1n_2}{\min\{n_1,n_2\} - n_1n_2}    \qquad \hbox{when} \quad  p \ge n_1n_2
\]
and 
\[ 
     \frac{p-n_1n_2}{n_1n_2 -\max\{0, n_1+n_2 - 1\}} \qquad \hbox{when} \quad p < n_1n_2 \ .
\]
  The probability tables in (\ref{second and third tables}) thus have $\kappa = -1$ and $+1$.  
The occupations of the two spin orbitals that are tabulated in (\ref{first table}) have a small correlation coefficient $\rho \approx -0.08$, but their marginal-normalized correlation $\kappa \approx -0.68$.  This value of $\kappa$ indicates that, {\it for two spin-orbitals whose occupation probabilities happen to be $0.9288$ and $0.0010$}, the occupations of these two spin orbitals are markedly anti-correlated.  

The value of $\kappa$ is very sensitive to the precision of the occupation probabilities.  Had we rounded the occupation probabilities to $5$ decimal places instead of $4$, $\kappa$ would have come out to be $-0.66$ instead of $-0.68$.    Fortunately, we do not require too much precision in $\kappa$, as we usually round it to one decimal place.

\end{document}